\documentstyle[12pt]{article}
\begin{document}
\begin{titlepage}

\hfill{UM-P-94/19}

\hfill{OZ-94/6}

\vskip 1 cm

\centerline{\Large \bf Possible reason why leptons are lighter
than quarks}

\vskip 1.5 cm

\centerline{\large
R. R. Volkas\footnote{U6409503@hermes.ucs.unimelb.edu.au}}

\vskip 5 mm

\noindent
\centerline{\it Research Centre for High Energy Physics}
\centerline{\it School of Physics}
\centerline{\it University of Melbourne}
\centerline{\it Parkville 3052 Australia}

\vskip 1.5 cm

\centerline{Abstract}

\noindent
The minimal model of spontaneously broken leptonic colour and discrete
quark-lepton symmetry predicts that charged leptons have the same
masses as their partner
charge $+2/3$ quarks up to small radiative corrections. By
invoking a different pattern of symmetry breaking, a similar model can be
constructed with the structural feature that charged leptons have to
be lighter than their partner quarks because of mixing between leptonic
colours, provided mixing between generations is not too strong.
As well as furnishing a new model-building tool,
this is phenomenologically interesting because some of the
new physics responsible for the quark-lepton mass hierarchy could
exist on scales as low as several hundred GeV.

\end{titlepage}

The patterns evident in the mass and mixing angle spectrum of quarks and
leptons continue to challenge us to provide an explanation. One may
broadly categorise these patterns as hierarchies between
generations, between weak-doublet partners and between quarks and leptons.
We do not know if these three sub-problems can be solved separately, or
if an all-encompassing explanation is necessary. In this paper I will
introduce a novel suggestion for why the charged lepton is less
massive than the charge $+2/3$ quark for a given generation. An
analysis of its explanatory success will lead
us to discuss how the quark-lepton hierarchy problem might be connected
with other hierarchy problems.

The obvious place to look for a reason for quarks to be heavier than
leptons is in the dynamics of colour. Is there any reason why coloured
fermions in a generation should be more massive than colourless
fermions? There is a well-known answer to this question in the context
of ultra-high--scale unification theories: If quarks and leptons have
similar or equal running masses in the $10^{15}$ GeV to Planck mass
regime, then gluonic interactions affect the running to lower energies
so as to raise quark masses by roughly the correct amount
relative to lepton masses \cite{runningmass}.
However, evolution through thirteen orders of magnitude or more in
energy is required since the masses run only logarithmically. Although
this is an interesting observation, it has the observational
disadvantage that the new physics of mass generation would be difficult
to test properly. Is there a way to understand the quark-lepton mass
hierarchy through new colour physics at much lower energy scales?

One likely avenue is through a spontaneously broken colour group for {\it
leptons} and discrete quark-lepton (q-$\ell$) symmetry
\cite{footlew}. These ideas have been
pursued for the last few years \cite{qlsym}.
The original motivation for them was to
connect the quantum numbers of quarks and leptons by new symmetries
that could be spontaneously broken at a relatively low scale such as 1
TeV. However, increasing symmetry
beyond the SU(3)$\otimes$SU(2)$\otimes$U(1) of the Standard Model (SM)
can also relate parameters such as coupling constants that were previously
unrelated. Indeed, it was immediately
noticed that in the minimal model discrete q-$\ell$ symmetry enforced the
tree-level mass relations
\begin{equation}
m_{e,\mu,\tau} = m_{u,c,t}\quad {\rm and}\quad m^{\rm
Dirac}_{\nu_e,\nu_\mu,\nu_\tau} = m_{d,s,b}.
\label{massrel}
\end{equation}
The most constructive way to view the phenomenologically
unacceptable charged-lepton--up-quark equality is as a spring-board for
further pondering. Although it is unacceptable {\it per se}, we after
all ultimately do want a theory that will relate quark and lepton
masses. I will show how this equality can be transformed into an
explanation for why charged leptons are less massive than their up quark
partners. (The $m^{\rm Dirac}_{\nu} = m_d$ equality
is perfectly acceptable if one
uses the see-saw mechanism \cite{ord_seesaw} to
explain why the standard neutrinos have such tiny masses.)

We begin by supposing that the SM gauge group $G_{SM}$ is embedded within
the larger gauge group $G_{q\ell}$ where
\begin{equation}
G_{q\ell} \equiv {\rm SU}(3)_{\ell}\otimes {\rm SU}(3)_q\otimes{\rm
SU}(2)_L\otimes{\rm U}(1)_X,
\end{equation}
where SU(3)$_q$ is the usual quark colour group, SU(3)$_\ell$ is
leptonic colour and $X$ is an Abelian charge.
A fermionic generation is assigned to representations
of $G_{q\ell}$ in the following way:
\begin{eqnarray}
&Q_L \sim (1,3,2)(1/3),\qquad u_R \sim (1,3,1)(4/3),\qquad d_R \sim
(1,3,1)(-2/3),&\ \nonumber\\
&F_L \sim (3,1,2)(-1/3),\qquad E_R \sim (3,1,1)(-4/3),\qquad N_R \sim
(3,1,1)(2/3).&\
\end{eqnarray}
This pattern is anomaly-free, and it enables us to define
the discrete symmetry
\begin{equation}
Q_L \leftrightarrow F_L,\quad u_R \leftrightarrow E_R,\quad d_R
\leftrightarrow N_R,\quad
G^{\mu}_q \leftrightarrow G^{\mu}_{\ell},\quad W^{\mu} \leftrightarrow
W^{\mu},\quad C^{\mu} \leftrightarrow -C^{\mu},
\label{qlsym}
\end{equation}
between the quarks and the generalized lepton fields $F_L$, $E_R$ and
$N_R$, and between the various gauge boson multiplets ($G^{\mu}_q$ are
gluons, $G^{\mu}_{\ell}$ are leptonic colour gluons, $W^{\mu}$
are weak gauge bosons and $C^{\mu}$ is the gauge boson for $X$).

The standard leptons lie within
$F_L$, $E_R$ and $N_R$. Their precise identification depends on the
spontaneous symmetry breakdown pattern of SU(3)$_{\ell}\otimes$U(1)$_X$.
In the models investigated
hitherto, the breakdown
SU(3)$_{\ell}\otimes$U(1)$_X$ $\to$
SU(2)$'\otimes$U(1)$_Y$ was employed (see Refs.~\cite{footlew,qlsym} for
details). This led to standard leptons being located purely
in the $T_8 = -2$ component of the triplets, where $T_8$ $\equiv$ ${\rm
diag}(-2,1,1)$ is a generator of SU(3)$_{\ell}$.
As in the SM, one introduced a single electroweak
Higgs doublet $\phi \sim (1,1,2)(1)$. Because of the q-$\ell$ discrete
symmetry of Eq.~(\ref{qlsym}), its Yukawa interactions were constrained
to be
\begin{equation}
{\cal L}_{\rm Yuk} = {m_u \over u} (\overline{Q}_L u_R \phi^c +
\overline{F}_L E_R \phi)
 + {m_d \over u} (\overline{Q}_L d_R \phi +
\overline{F}_L N_R \phi^c) + {\rm H.c.}
\label{ewyuk}
\end{equation}
where $\phi^c \equiv i\tau_2\phi^*$.
Electroweak symmetry breakdown via $\langle\phi\rangle = (0,u)^T$
then produced the quark-lepton mass relations because of the discrete
q-$\ell$ symmetry (under which $\phi$ $\leftrightarrow$ $\phi^c$).

That standard leptons
possessed a unique leptonic colour was important in this derivation.
If standard leptons were a superposition of components of {\it different}
leptonic colour, then Eq.~(\ref{ewyuk}) would not necessarily produce
the mass relations of Eq.~(\ref{massrel}). The model I construct below
is based on this observation.

To proceed, we need to spontaneously break the SU(2)$'$ subgroup of
SU(3)$_{\ell}$ that was left exact hitherto.
We will use the same gauge group $G_{q\ell}$ and fermion
representation content as the usual q-$\ell$ symmetric model,
but a different Higgs sector.

The non-electroweak Yukawa Lagrangian is given by ${\cal L}'_{\rm Yuk}$
where
\begin{eqnarray}
{\cal L}'_{\rm Yuk} & = & h_1[\overline{(F_L)^c} F_L \chi +
\overline{(Q_L)^c} Q_L \chi'] + h_2 [\overline{(N_R)^c} E_R \chi +
\overline{(d_R)^c} u_R \chi']\nonumber\\
& + & h_3[\overline{(N_R)^c} N_R \xi + \overline{(d_R)^c} d_R \xi'] +
h_4 [\overline{(N_R)^c} E_R \Delta + \overline{(d_R)^c} u_R \Delta']
\nonumber\\
& + & h_5 [\overline{(F_L)^c} F_L \Delta + \overline{(Q_L)^c} Q_L \Delta']
+ {\rm H.c.}
\label{newyuk}
\end{eqnarray}
and the Higgs boson q-$\ell$ symmetry pairs are
\begin{eqnarray}
&\chi \sim (3,1,1)(2/3)\quad {\rm and}\quad \chi' \sim (1,3,1)(-2/3),&\
\nonumber\\
&\xi \sim (3,1,1)(-4/3)\quad {\rm and}\quad \xi' \sim (1,3,1)(4/3),&\
\nonumber\\
&\Delta \sim (\overline{6},1,1)(2/3)\quad {\rm and}\quad
\Delta' \sim (1,\overline{6},1)(-2/3).&\
\end{eqnarray}
The electroweak Higgs sector again contains one
electroweak doublet $\phi$, and the Yukawa Lagrangian is the same as
Eq.~(\ref{ewyuk}).

Spontaneous symmetry breaking proceeds in at least two stages. First,
the fields $\chi$, $\xi$ and $\Delta$ gain nonzero vacuum expectation
values (VEVs) to break both leptonic colour and the discrete q-$\ell$
symmetry, leaving electroweak SU(2)$_L\otimes$U(1)$_Y$ unbroken.
(The partner Higgs fields $\chi'$, $\xi'$ and $\Delta'$ must
of course have zero VEVs to keep standard colour exact.)
The non-standard fermions in the theory gain nonzero masses
via ${\cal L}'_{\rm Yuk}$ at this stage. The standard leptons (and
quarks) are defined to be those fermions that remain massless.
The electroweak gauge symmetry is then broken in the second stage
of symmetry breaking via the usual nonzero VEV for $\phi$. This also generates
nonzero masses for the standard leptons. For phenomenological reasons we
will require that $\langle\chi\rangle$, $\langle\xi\rangle$,
$\langle\Delta\rangle$ $\gg$ $\langle\phi\rangle$.

The VEVs of the leptonically coloured Higgs bosons are induced to
take the forms
\begin{equation}
\langle\chi\rangle = \left( \begin{array}{c} w \\ 0 \\ 0 \end{array}
\right),\quad
\langle\xi\rangle = \left( \begin{array}{c} 0 \\ v \\ 0 \end{array}
\right)\quad {\rm and}\quad
\langle\Delta\rangle = \left( \begin{array}{ccc}
0 & a & 0 \\ a & 0 & 0 \\ 0 & 0 & 0 \end{array} \right),
\label{vevs}
\end{equation}
where $\Delta$ is represented by a $3\times3$
symmetric matrix.\footnote{It is important to check that this VEV
pattern can be the minimum of the Higgs potential for a range of
parameters. The Higgs potential $V$ for this model is quite complicated,
and I will not write it down in this paper since I want to focus
on the issue of fermion mass. Ideally, one would like to write $V$ in
the sum-of-squares form $\Sigma\lambda_i|{\rm quadratic\ form}|^2$,
where the $\lambda_i$'s are chosen to be positive
and ``quadratic form'' is a quadratic
function of the Higgs fields. The global minimum of $V$ is then
obtained by simply making each quadratic form zero.
My analysis shows
that most of the terms in $V$ can be written in this manner in such
a way that the required alignment of nonzero VEVs ensues. There are a
few terms that I have not succeeded in writing thus, so the required
region of parameter space may force the
coefficients of these recalcitrant terms to be somewhat smaller than
the $\lambda_i$'s. If these coefficients are zero, then it turns out
there is an
unwanted global U(1) symmetry in $V$ and an unwanted pseudo-Goldstone
boson is produced. A rigorous analysis would need to show that this
potentially light boson is made sufficiently heavy when the
recalcitrant terms are switched on. This issue is beyond the scope of
this paper.}
This VEV pattern induces the breakdown
\begin{equation}
{\rm SU}(3)_{\ell}\otimes{\rm SU}(3)_q\otimes{\rm SU}(2)_L\otimes{\rm
U}(1)_X \to {\rm SU}(3)_q\otimes{\rm SU}(2)_L\otimes{\rm U}(1)_Y \equiv
G_{SM},
\end{equation}
where weak hypercharge $Y$ is given by
\begin{equation}
Y = X + {T_8 \over 3} + T_3,
\label{Y}
\end{equation}
with $T_3$ being the diagonal generator ${\rm diag}(0,1,-1)$ of leptonic
colour. The major difference between this model
and the usual q-$\ell$ symmetric model is the presence of $T_3$
in the formula for $Y$.

Using Eq.~(\ref{Y}) we see that the leptonic colour components of the
generalized leptons have weak hypercharges given by
\begin{eqnarray}
&Y(F_L) = Y\left( \begin{array}{c} \ell_L \\ (f_R)^c \\ f_L \end{array}
\right) = \left( \begin{array}{c} -1 \\ +1 \\ -1 \end{array} \right),\quad
Y(E_R) = Y\left( \begin{array}{c} e_{1R} \\ \nu_{2R} \\ e_{3R} \end{array}
\right) = \left( \begin{array}{c} -2 \\ 0 \\ -2 \end{array}
\right),&\ \nonumber\\
&Y(N_R) = Y\left( \begin{array}{c} \nu_{1R} \\ (e_{2L})^c \\ \nu_{3R}
\end{array} \right) = \left( \begin{array}{c} 0 \\ +2 \\ 0 \end{array}
\right),&\
\label{Yforcomps}
\end{eqnarray}
where we have used a suggestive notation for the colour components of
$F_L$, $E_R$ and $N_R$. As a further piece of notation,
let the weak-isospin components of the colours of $F_L$ be given by
\begin{equation}
\ell_L = \left( \begin{array}{c} \nu_L \\ e_L \end{array} \right),\quad
(f_R)^c =
\left( \begin{array}{c} (\epsilon_R)^c \\ (n_R)^c \end{array} \right)
\quad {\rm and}\quad
f_L = \left( \begin{array}{c} n_L \\ \epsilon_L \end{array} \right).
\end{equation}

Equation (\ref{Yforcomps}) reveals that
the generalized leptons contain, per generation,
(i) the standard leptons plus a right-handed neutrino,
(ii) mirror or vector-like pairs of $\ell_L$-like and $e_R$-like
states and (iii) two additional $\nu_R$-like particles.
After the first stage of symmetry breakdown the vector-like pairs form
massive Dirac fermions, while the $\nu_R$-like states all become
massive. The standard $\ell_L$ and $e_R$ leptons are defined to be
the remaining massless states.

By inputting the VEVs $\langle\chi\rangle$, $\langle\xi\rangle$,
$\langle\Delta\rangle$ and $\langle\phi\rangle$
into the Yukawa Lagrangian of Eqs.~(\ref{newyuk}) and (\ref{ewyuk})
we find the charged lepton mass matrix
to be given by
\begin{equation}
{\cal L}^{\rm ch\ lept}_{\rm Yuk} =
\left( \begin{array}{ccc} \overline{e}_L & \overline{e}_{2L} &
\overline{\epsilon}_L \end{array} \right)
\left( \begin{array}{ccc}
m_u & 0 & M_5^{\dagger} \\ M_4 & M_2 & m_d^T \\
0 & m_u & M_1 \end{array} \right)
\left( \begin{array}{c} e_{1R} \\ e_{3R} \\ \epsilon_R \end{array}
\right) + {\rm H.c.}
\label{emass}
\end{equation}
where
\begin{equation}
M_5 \equiv (h_5 - h_5^T) a,\quad
M_4 \equiv h_4 a,\quad M_2 \equiv h_2 w\quad {\rm and}\quad M_1 \equiv
(h_1 + h_1^T) w.
\end{equation}
The terms $m_u$, $m_d$, $M_{1,2,4,5}$ are all $3 \times 3$ matrices in
generation space. The matrix $M_5$ is antisymmetric in generation space
and thus plays no role in a one-generation toy version of the model.
Let us call the full mass matrix in Eq.~(\ref{emass}) $M_{-}$.

To get a feel for what this mass matrix does, let us turn off the
generation structure for the moment (which means that $M_5 = 0$).
In the absence of the electroweak
contributions $m_u$ and $m_d$ we see that:

\vskip 2 mm

\noindent
(1) The $Q=-1$ field $e_L$ is massless and thus identified as
the standard left-handed electron. The other charged
members $\epsilon_{L,R}$
of $F_L$ form the left- and right-handed components
of a Dirac fermion of mass $M_1$. (Actually the whole
weak-doublet $f = f_L + f_R$
of Dirac fermions has mass $M_1$.)

\vskip 2 mm

\noindent
(2) The fields $\epsilon'_L \equiv e_{2L}$ and $\epsilon'_R \equiv
\sin\phi\ e_{3R} + \cos\phi\ e_{1R}$ form a $Q=-1$ Dirac fermion
$\epsilon'$ of mass $\sqrt{M^2_2 + M^2_4}$,
where $\tan\phi = M_4/M_2$.
The right-handed field orthogonal to $\epsilon'_R$ is
massless and thus identified as the standard right-handed electron: $e_R
\equiv \cos\phi\ e_{3R} - \sin\phi\ e_{1R}$.

\vskip 2 mm

When the electroweak terms $m_u$ and $m_d$ are switched on, $e_L$ and
$e_R$ are connected by a diagonal mass and they also mix with the heavy
exotic electron-like states $\epsilon$ and $\epsilon'$. The
mass of the physical standard electron is the
magnitude of the smallest eigenvalue of $M_{-}$.
Continuing to ignore generation
structure, we see that ${\rm det}M_{-} = m_u (M_1 M_2 - m_u m_d)$. From
$\langle\chi\rangle$, $\langle\xi\rangle$, $\langle\Delta\rangle$ $\gg$
$\langle\phi\rangle$
we expect that $M_{1,2,4} \gg m_{u,d}$ so that
${\rm det}M_{-} \simeq m_u M_1 M_2$. To zeroth order in $m_{u,d}$ the
large eigenvalues are still $M_1$ and $\sqrt{M^2_2 + M^2_4}$,
so see we that
\begin{equation}
{\rm smallest\ eigenvalue} \equiv m_e \simeq m_u \cos\phi \le m_u,
\label{result}
\end{equation}
where
\begin{equation}
\cos\phi \equiv {M_2 \over \sqrt{M^2_2 + M^2_4}}.
\label{cosphi}
\end{equation}
This equation illustrates the central result of this paper: {\it
Mixing between electron-like states of different leptonic colour
lowers the electron mass from that of its q-$\ell$ symmetric partner the
up quark}.\footnote{This type of result was first
explicitly calculated by M. de Jonge \cite{deJonge} in the context
of a quark-lepton symmetric model with a Higgs sector different from
the one I use here. The primary characteristic of the Higgs sector
used in Ref.~\cite{deJonge} was the non-minimal
combination of two $\chi$-type leptonic triplets.}

In the multi-generation real world,
each of the ratios $m_e/m_u$, $m_{\mu}/m_c$ and
$m_{\tau}/m_t$ can be separately adjusted to fit the measurements. This
can be trivially seen by supposing we have three generations but no
inter-generational mixing. In this case, the charged-lepton of each
generation is lighter than the corresponding up-quark. Furthermore,
each generation can have different values for
their corresponding $M_2$ and $M_4$ masses, so the corresponding
values for $\cos\phi$ can be different. Not unexpectedly,
inter-generational mixing complicates this picture. If generations mix
strongly, then it is no longer true that all charged lepton
masses are {\it necessarily} lower than their partner
up-quarks. The precise meaning of ``partner'' is of course
unclear in the presence of inter-generational mixing. I will discuss
this issue in more depth after I conclude the technical exposition.

The neutrino sector before electroweak symmetry breakdown splits into
massless left-handed neutrinos $\nu_L$,
massive fermions $n \sim n_L + n_R$ which are degenerate with the
$\epsilon$'s, plus a right-sector mass matrix given by
\begin{equation}
{\cal L}^{\rm neut}_{\rm mass} =
\left( \begin{array}{ccc} \overline{(\nu_{1R})^c} &
\overline{(\nu_{2R})^c} & \overline{(\nu_{3R})^c} \end{array} \right)
{1 \over 2} \left( \begin{array}{ccc} 0 & M_4 & M_3 \\
M_4^T & 0 & -M_2^T \\ -M_3 & -M_2 & 0 \end{array} \right)
\left( \begin{array}{c} \nu_{1R} \\ \nu_{2R} \\ \nu_{3R} \end{array}
\right) + {\rm H.c.}
\end{equation}
where $M_3 \equiv (h_3^T-h_3)v$ and $M_{2,3,4}$ are $3\times3$ matrices
in generation space.
Diagonalization of the whole right-sector neutrino mass matrix
yields all eigenvalues as nonzero and of order $M_i$,
so the usual see-saw mechanism \cite{ord_seesaw} for
neutrinos will ensue when electroweak symmetry breakdown occurs (note
that $M_3 \neq 0$ is essential).

The eight gauge bosons of leptonic colour acquire masses of the order
of $g_s(\Lambda)\Lambda$ where $g_s$ is the strong coupling constant and
$\Lambda$ $\sim$ $\langle\chi\rangle$,
$\langle\xi\rangle$ and $\langle\Delta\rangle$. Neutral current and
other phenomenology will typically require that these gauge bosons
are heavier than about 1 TeV, so that $\Lambda$ $\sim$
$\langle\chi\rangle$,
$\langle\xi\rangle$, $\langle\Delta\rangle >$ 1 TeV.
Note that the $\cos\phi$ suppression factor can be large even if the
leptonic colour breaking scale is much higher than a few TeV. This
is because the mixing between electron-like states of different leptonic
colour is controlled by $M_4$ and thus it increases with $\Lambda$.

Let us now evaluate the successes and failures of the above scenario:

\vskip 2mm

\noindent
(i) We have succeeded in constructing a quark-lepton symmetric
model that has both a minimal electroweak Higgs sector and acceptable
quark-lepton mass relations. By contrast, in the usual
q-$\ell$ symmetric model one
evades the relations in Eq.~(\ref{massrel}) by postulating two
electroweak Higgs doublets rather than one \cite{levin}.

\vskip 2 mm

\noindent
(ii) But the most important achievement is
the fact that charged leptons are forced to be less massive
than up-quarks by a {\it structural ingredient of the model}. This
provides us with an interesting new technique in model-building, and
is the main point of this paper. My
mechanism is closely related in spirit to the
see-saw mechanism \cite{ord_seesaw} and the
universal see-saw mechanism \cite{univ_seesaw}.
The former is a way of using fermion mixing
to understand why neutrinos are much lighter than any other fermion,
while the latter is a way of employing fermion mixing to
understand why fermions are generally much lighter than the
electroweak scale. My mechanism, on the other hand, is a way to employ
fermion mixing to understand why charged leptons are lighter than up
quarks. Furthermore, it is a low-energy (or potentially
low-energy) alternative to the running mass idea
alluded to in the introductory paragraphs. I stress also
that the mechanism itself is almost certainly of more interest than the
specific model I have chosen by way of illustration here.
(This is after all
also true of the see-saw and universal see-saw mechanisms.) For
instance, there are non-minimal
q-$\ell$ symmetric models \cite{nonminimal_ql} that have $m_e
= m_d$ rather than $m_e=m_u$ which can also probably be modified to
incorporate my mechanism. This could be of great interest since the
charged-lepton--down-quark hierarchy is less severe than the
charged-lepton--up-quark hierarchy for the second and third generations,
and thus it might be easier to explain.

\vskip 2 mm

\noindent
(iii) The specific model examined here
easily incorporates both the see-saw
mechanism for neutrinos and my new mechanism in a reasonably coherent
theoretical structure.

\vskip 2 mm

\noindent
(iv) The model, however, fails to account for the
quark-lepton mass hierarchy in quantitative detail.
In the one-generation case of Eqs.~(\ref{result}) and (\ref{cosphi})
we require $M_4$ to be significantly larger than $M_2$ in order to
reproduce any of $m_e/m_u$, $m_{\mu}/m_c$ or $m_{\tau}/m_t$. It is
interesting that $M_4$ is proportional to the sextet Higgs VEV while
$M_2$ is proportional to a triplet Higgs VEV. This indicates that the
quark-lepton mass hierarchy might be related to a VEV hierarchy and we
would have to search for a fundamental reason for the sextet Higgs to
have a larger VEV than the triplet Higgs. But we also note that
such a VEV hierarchy
is not enough since the additional hierarchy $m_e/m_u$ $>$ $m_{\mu}/m_c$
$>$ $m_{\tau}/m_t$ can only be incorporated by adjusting Yukawa coupling
constants.

However, I argue that
it is inappropriate to demand of the present
model that it explain the quark-lepton hierarchy in this much detail, since
it does not seek to address any of the other hierarchy sub-problems.
This echoes the point I made in the opening
paragraph that the various sub-problems within the global conundrum of
fermion mass may well interconnect in non-trivial ways. It
would be very surprising in my view if a theory {\it perfectly}
explained
one type of hierarchy in the quark-lepton sector but left the others
accomodated but unexplained. I have deliberately made no attempt to
address the generation, mixing angle and down-up quark hierarchies in
the present model, because I wanted to discuss the
quark/lepton mass hierarchy sub-problem in an unencumbered context.
However, ultimately it is probably true that only those models that seek
to explain the whole lot can adequately explain any one specific
hierarchy.

This of course relates to what becomes of my mechanism when mixing
between generations is switched on. As stated previously, if generations
mix together strongly then it is no longer necessary for all charged-leptons
to be lighter than their ``partner'' up-quarks. However,
the only known example of fermion mixing in the real world is
parameterised by the approximately diagonal Kobayashi-Maskawa matrix.
This suggests that nature may suppress mixing between generations in
general. On the other hand, this is by no means inevitable or even desirable.
The way ahead is unclear because I can make arguments both for
and against a strong correlation between the quark/lepton and
inter-generational hierarchy subproblems:

\vskip 2 mm

\noindent
{\it The case for.} One can speculate that the $M_2$ and $M_4$
parameters in $\cos\phi$ perhaps should come from different
generations so that $M_4 > M_2$ for the same general reason that, say,
$m_c > m_u$. This would require a special pattern
of generation mixing that might be due, for example, to
a very particular horizontal symmetry. It would also remove the
motivation for the proposed
hierarchy between the sextet and triplet Higgs boson VEVs.

\vskip 2 mm

\noindent
{\it The case against.}
We know from the Kobayashi-Maskawa matrix that the only observed mixing
between generations is suppressed. We also believe the
structural feature that
charged-leptons are necessarily lighter than up-quarks to be an elegant
feature of a one-generation toy model version of my theory. It is
therefore tempting to combine these observations and so require of an
extension of the present theory (that non-trivially incorporates
generations) that it necessarily prohibit strong mixing between
generations. The role of the matrix $M_5$ is of particular
interest in this regard because it is purely non-diagonal, and thus it
may be forbidden by this hypothetical extension.

\vskip 2 mm

These are difficult issues, and they perhaps go to the core of why the
fermion mass problem has proven so intractable. One can either
attempt a grand explanation of all features of the problem at once, or
one can attack sub-problems individually and then try to synthesize
the disparate elements later. The grand approach is probably too
ambitious, while in the piecemeal approach one is bound to be
dissatisfied with the explanatory powers of the individual pieces.

My model follows the piecemeal philosophy, and so one
has to be perfectly clear
about the scope of the exercise in order to not make inappropriate
demands. Our purpose here was to address some ``fine-structure''
within a generation and the model I propose broadly
achieves this aim. How this
picture should be extended to incorporate (a) multiple generations in a
profound way (horizontal symmetry? compositeness?), and (b) the up-quark
to down-quark hierarchy, is a task for the future.
(Of course, these types of observations can also be made
about the see-saw and universal see-saw mechanisms. Neither has anything
to say about generations, mixing angles or the weak-isospin hierarchy.)

\newpage

\centerline{\bf Acknowledgements}

I would like to thank Martin de Jonge for many discussions on
completely broken leptonic colour, Robert Foot for reading a draft
version and Henry Lew for comments. This work was supported by the
Australian Research Council and the University of Melbourne.

\end{document}